\documentclass[twocolumn,amsmath,amssymb]{revtex4}
\usepackage{amstext}
\usepackage{amsmath}
\usepackage{amssymb}
\usepackage{graphicx}
\usepackage{subfigure}

\usepackage{color} 

\begin{document}
\title{On the phase behavior of hard aspherical particles}
 
\author{William L. Miller} 
\author{Angelo Cacciuto}
\email{ac2822@columbia.edu}
\address{Department of Chemistry, Columbia University, 3000 Broadway,\\New York, New York 10027}

\begin{abstract}
We use numerical simulations to understand how random deviations from the ideal spherical shape affect the ability of hard particles to form 
{fcc} crystalline structures. Using a system of hard spheres as a reference, 
we determine the fluid-solid coexistence pressures of both shape-polydisperse and monodisperse systems of aspherical {hard particles}.
We find that when {particles} are sufficiently isotropic,  the coexistence  pressure can be predicted from 
a linear relation involving the product of two simple geometric parameters characterizing the asphericity  of the particles.
Finally, our results allow us to gain direct insight into the crystallizability limits of these systems by rationalizing 
empirical data obtained for analogous monodisperse systems~\cite{disorder1}.

\end{abstract}
\maketitle

\section{INTRODUCTION}

Understanding how objects pack or can be designed to tile the three dimensional space is a fundamental optimization problem that has important practical applications that
range from the macroscopic, such as the efficient storage of grains,  to the microscopic: the fabrication of band-gap photonic materials.
Mathematicians have been intrigued by such problems for centuries; namely since Kepler's 1611 essay {\it On the Six-cornered Snowflakes}, but it wasn't 
until recently, with the advent of nanotechnology and the explosion of molecular biology, that problems of packing and self-assembly of nanoscopic components 
gained   tremendous traction in the broader scientific community.   
Recent advances in synthesis of nanoparticles~\cite{DeVries,Schnablegger,Hong,Weller,Hobbie,weitz,pine,mitragotri,kraft} allowed for unprecedented control over the 
 shape and surface chemistry of colloidal particles, thus providing an unlimited number of building blocks whose 
 spontaneous aggregation could lead to the formation of an unprecedented variety of structures with potentially novel functional, mechanical, and optical properties.
 Unlike most of the work on particle crystallization and self-assembly that in the last decade has focused on 
 monodisperse~\cite{frenkel1} or polydisperse~\cite{frenkel2,zaccarelli} systems of spherical or regularly-shaped particles  (see also~\cite{torquato2,glotzer,chandler,geissler,cacciuto,torquato,frenkel,glotzer2,glotzer3,glotzer4,esco,weitz} and references therein), surprisingly little or nothing has been done
 theoretically to understand the packing of irregularly shaped particles. 
Indeed, there are several important cases in which the shape of the single components cannot be tailored at will,
 yet, an efficient packing, or an understanding of the physical properties of these densely compressed systems, is highly desirable.
{An example of an outstanding problem in this category is the storage of grains~\cite{deGennes}.}
  
Here we seek to gain insight into the crystallization of {hard aspherical} particles. 
 We want to understand how  particle geometric features can be related to their ability to orderly pack into three dimensional periodic structures,
 and especially  to identify under what conditions they cease to do so. 
 We have recently reported~\cite{disorder1}  that it is possible to empirically relate particle geometry to crystallizability (intended as the tendency of a component to crystallize) 
by using two simple geometric parameters. The first is the particle asphericity  $A$, defined in terms of the 
surface to volume ratio of a particle $\alpha_p=A_p/V_p$ with respect to that of a sphere of diameter $\sigma$, $\alpha_s=6/\sigma$, as
$A =1-\alpha_s/\alpha_p.$  The second parameter, $q$, is related to the  orientational symmetry  of the particle.
It is used to describe the  asphericity of random walks~\cite{rudnick}, and it is obtained by combining invariants of the particle inertia tensor $I_{ij}$  as
$q=\sum_{i< j} (\lambda_i^2-\lambda_j^2)^2/\left(\sum_i \lambda_i^2\right)^2$, where $\lambda_i$, with $i=1, 2, 3$, is an eigenvalue of $I_{ij}$.
 
In this paper we try to rationalize those empirical results by computing how random perturbations from the ideal spherical 
shape affect the fluid-solid coexistence pressure of  monodisperse and shape-polydisperse systems of hard aspherical particles. 

\section{METHOD}

In order to generate a  statistical ensemble of {hard} aspherical particles, we developed a simple model~\cite{disorder1} that guarantees a certain degree of control over
the particle shape. Each particle is built by setting the center of $N_b$ ($4 \leq N_b \leq 12$) spheres of diameter $\sigma$ at random 
positions on the surface a {sphere} of diameter $\sigma_0<\sigma${, allowed to overlap freely}.
Polydisperse systems of such aspherical particles were generated by choosing specific values of $N_b$ and $\sigma_0$ and allowing each particle in the system to arise from a different random collection of sphere positions.
Deviations from the spherical shape can be conveniently controlled by varying $\sigma_0$ and $N_b$.

{All dimensions of the resulting particles are then scaled so that the volume $V_p$ of each particle is equal to that of a sphere with radius $\sigma$, $V_s = \frac{1}{6}\pi \sigma^2$.
The resulting sphere diameters after this rescaling are denoted $\sigma_R$.}
For  $\sigma_0=0$ one recovers the spherical limit, and as  $\sigma_0$ increases, particles develop larger and larger shape distortions.
In a similar fashion, large values of $N_b$ result in a bumpy but overall isotropic particle, whereas small values of $N_b$ tend to generate very anisotropic shapes.
Any two particles $i$ and $j$ interact via a hard repulsive potential defined as
\begin{equation}
U_{ij}=
	\begin{cases}
		0 & {\textrm{if }} |r_s-r_t|>\sigma_R \,\,\,\,\,\,\forall s\in i \,\,,\,\, \forall t\in j \cr
		\infty &\text{otherwise}\cr
	\end{cases}
\end{equation}
where $s$ and $t$ run over all spheres of rescaled diameter $\sigma_R$ constituting particle $i$ and particle $j$ respectively.
Experimental realizations of colloidal particles similar to ours could be generated using the approach described in reference~\cite{weitz,pine,mitragotri} 
to create nonspherical particles with tunable shapes.  

In order to determine the fluid-solid coexistence pressures for both polydisperse and monodisperse systems of aspherical particle, we used the method of direct fluid-solid coexistence simulation described in~\cite{noya}.
1024 {hard} particles were placed in an FCC crystal lattice (at volume density $\rho_s \simeq 0.545$, the hard sphere crystal coexistence density~\cite{HScoex}) centered in a box of dimensions $L_x \times L_y \times L_z$.
An FCC crystal was chosen based upon our previous work, which indicated that when monodisperse aspherical systems formed crystals, they tended to be (apart from a few exceptions) FCC; we have found no evidence supporting the use of any other crystal geometry.
The dimensions of the box were chosen such that $L_x$ and $L_y$ were just large enough to accomodate the FCC crystal, and $L_z$ was roughly four times larger.
The crystal lattice was chosen such that the extension of the crystal in the $z$-direction was roughly twice that in the $x$- and $y$-directions, in order to increase the separation between the two fluid-solid interfaces in the system.
This crystal lattice was placed into equilibrium with a fluid of 1024 particles at hard sphere fluid coexistence volume density $\rho_l \simeq 0.495$.
The fluid and crystal were both briefly allowed to relax in order to relieve any overlap introduced by the ``bumpiness" of the particles and to allow the fluid to come fully into contact with the solid interface.

Monte Carlo simulations were then run in the $NPT$ ensemble, where the three box dimensions $L_x$, $L_y$, and $L_z$ were allowed to fluctuate independently.

The number of crystalline particles $N_X$ in the system was monitored as a function of Monte Carlo step; this quantity was determined using the standard spherical-harmonics based bond order parameter $q_6$~\cite{q61,q62}.
Direct simulation is known to have a few caveats: slow equilibration, non-insignificant finite size effects,  dependence of the surface free energy on the specific face the crystal exposes to the fluid~\cite{noya}.
Nevertheless, for this specific system, we find this direct  method to be  more reliable than the two-step thermodynamic integration scheme described in~\cite{dijkstra}. 
  
\begin{figure}
	\includegraphics[width=0.4\textwidth]{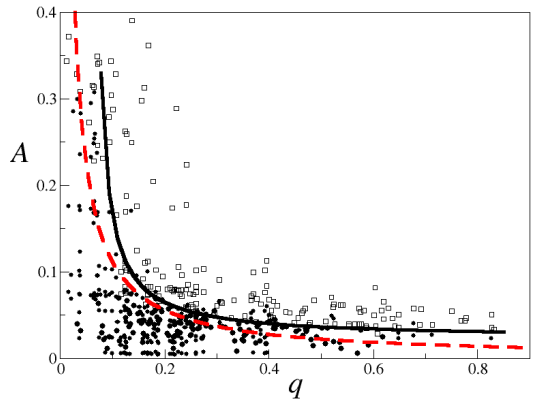}
	\caption{Crystallizability of monodisperse {hard} aspherical particles characterized in terms of two shape parameters $A$ and $q$. Filled circles indicate particles that easily crystallized, while open squares indicate particles that did not.  {The solid line is a hand-constructed fit to the crystallizability limit;} the dashed line represents a prediction of the crystallizability limit from the current work; see below.  Figure adapted from~\cite{disorder1}.}\label{mono1}
\end{figure}
Our empirical results on the crystallizability of   systems of monodisperse {hard} aspherical particles ~\cite{disorder1}, obtained by slowly compressing an ensemble of 487 different shape realizations {(see \cite{disorder1} for details)}
 are summarized in  Fig.~\ref{mono1} and indicate the existence of a clear boundary between particles that crystallize and particles that do not.
  A roughly inverse relationship is clearly evident; particles with large $A$ must have very small $q$ in order to have
a hope of crystallization, and vice-versa. This result provides a very useful way of predicting whether a particular particle shape can pack into a
crystalline structure by simply measuring the experimentally accessible $A$ and $q$. 
As this  model is intended to describe randomly shaped particles, the diagram does not include the results for particles designed with very specific shapes such as rods,
plates or regular polyhedric geometries that are known to crystallize. 
These particular cases would generate sharp peaks around specific values of $q$, and are purposefully excluded from this study.
 The solid line is  a guide to the eye and has the functional form $A(q) = 0.023 + 1/(170q-10)$. 
\begin{figure}
	\includegraphics[width=0.4\textwidth]{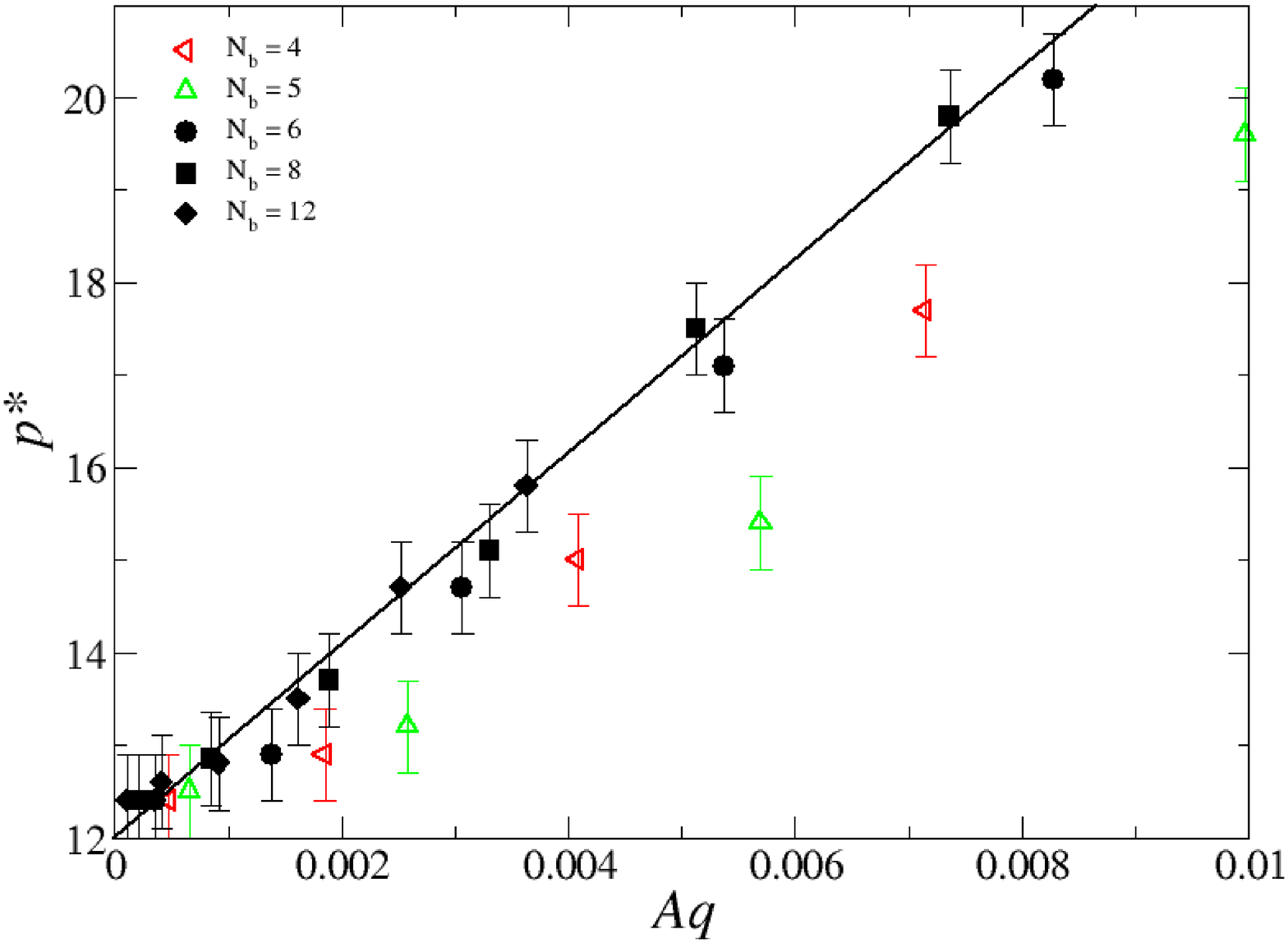}
	\put(-75,18){\includegraphics[width=0.135\textwidth]{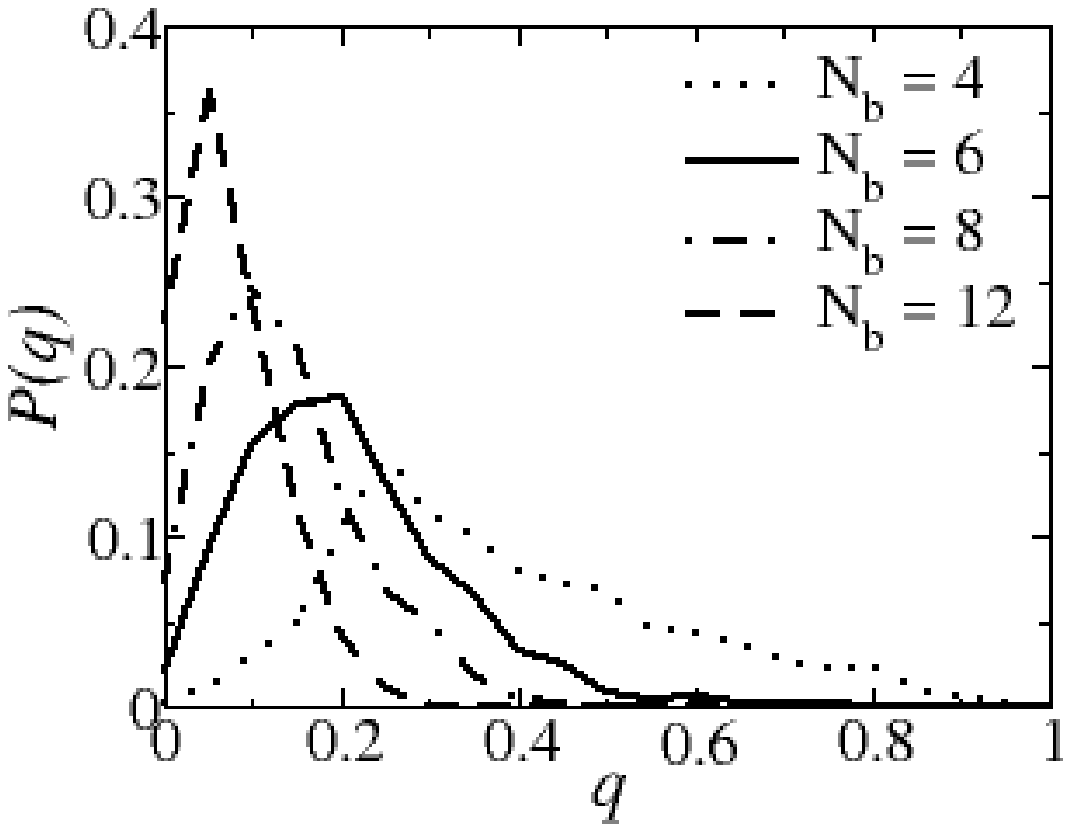}}
	\caption{Coexistence pressure vs. $Aq$ for shape-polydisperse systems of hard aspherical particles. The solid line is a fit to the data for $N_b\geq 6$.  Inset: Distribution of $q$ values for polydisperse systems with $N_b$ = $4$, $6$, $8$, and $12$.  Recall that $q$ is independent of $\sigma_0$.  The distribution is significantly larger and broader for $N_b = 4$ than for the larger values of $N_b$.}\label{poly}
\end{figure}

\section{POLYDISPERSE SYSTEMS}

The question remains as to why the systems behave this way; this is the focus of the current work.
To get a better understanding of the crystallizability limit,
we compute the coexistence pressure of systems of shape-polydisperse {hard} particles as a function of the geometric parameters $A$ and $q$ via the direct fluid-solid simulation scheme described above.
Incorporating shape-polydispersity in these systems is however not a trivial matter and requires some discussion.  
Unlike the common notion of polydispersity of spherical particles  for which any particle size can be indifferently used as a reference,  each particle shape whose values of $A$ and $q$ lead to a point below the crystallizability boundary shown in Fig.~\ref{mono1} could be used as a reference shape for this study. The issue is that the phase behavior could be very much dependent on
the specific choice of $A$ and $q$, making the problem intractable. We therefore adopted a pragmatic approach to describing shape-polydispersity  that has a natural experimental 
counterpart~\cite{weitz}. 
Namely, we  consider the distributions of $q$s and $A$s arising when constructing our particle using a given number of spheres $N_b$ and a given value of $\sigma_0$.
Particles  with large values of $N_b$ generate narrow distributions shifted towards small values of $q$, while small values of $N_b$ ($N_b\geq 4$) result in wide distributions peaked over large values of $q$ {(see Figure~\ref{poly}(a))}.
When $\sigma_0\rightarrow 0$ one recovers the  hard sphere limit for any value of $N_b$, and as $\sigma_0$ increases,  larger and larger values of $A$ 
will be sampled (without affecting the $q$ distributions) until  approaching the crystallizability boundary. 

We systematically analyzed {hard} particles constructed with $N_b =$4, 5, 6, 8, and 12, and values of  
$\sigma_0 \in[0.05,0.30]$.

All data points have been computed using two different random sets of particle geometries from the given distribution.
The difference between the results from the two independent sets is smaller than the error bar associated with the numerical scheme.
Figure~\ref{poly} shows our results. 
Remarkably, we find that the coexistence pressures, $p^*(A,q)$, between fluid and solid for values of $N_b\geq6$
can be made to collapse into a simple linear dependence on $Aq$ which  can be fitted to  $p^* (A,q)= 991Aq + 12$;
note that the y-intercept is near the hard sphere coexistence pressure of $p^*_{\rm HS}\simeq  11.7$~\cite{HScoex}, as it should be.
Our results thus suggest that $p^*(A,q)$ can be expressed in terms of a Taylor expansion in the quantity $Aq$: $\frac{p^*(A,q)-p^*_{\rm HS}}{p^*_{\rm HS}} = \alpha Aq + ...$.
Note, however, that significant but systematic deviations from the linear fit are visible for systems of particles 
obtained with  $N_b = 4$ and $N_b = 5$. 
These values of $N_b$ correspond to particles with  relatively large values of $q$ and rather broad distributions (see Figure~\ref{poly} inset), suggesting that higher order terms in the expansion may be necessary. Furthermore, for such wide distributions 
fractionation  between isotropic and anisotropic particles may become an important factor.
\begin{figure}
	\includegraphics[width=0.4\textwidth]{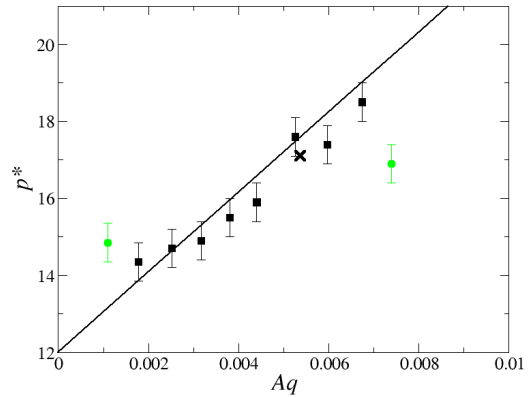}
	\caption{Coexistence pressure vs. $Aq$ for monodisperse systems of hard aspherical particles with $N_b = 6$, $\sigma_0 = 0.2$.  The line is a fit obtained from the data presented in Figure~\ref{poly} for polydisperse systems with $N_b = 6$, $8$, and $12$.  The cross indicates the position of the polydisperse system with $N_b = 6$, $\sigma_0 = 0.2$.
Points at the extremes of the $q$ distribution are shaded with a lighter color.}\label{mono2}
\end{figure}

\section{APPLICATION TO MONODISPERSE SYSTEMS}

The existence of a master curve for the coexistence pressure of significantly different particle distributions (provided $N_b\geq6$), both in terms of their average value  and their width, is quite remarkable, and provides a very compact and elegant way of expressing the equilibrium properties of aspherical systems, thus significantly reducing the number of variables required to describe these complex systems.  
The results from the above study are very robust.
In fact, we next considered 10 monodisperse systems generated by a single set of parameters: $N_b = 6$, $\sigma_0 = 0.2$.
Each of these systems is characterized, by definition, by an infinitely sharp $P(q)$, 
and individual particle geometries were selected to mostly cover the range of $Aq$ values explored above, while keeping $A$ roughly constant because for a given $N_b$ and $\sigma_0$, the distribution of $A$ is quite narrow. 
A plot of these data, shown in Figure~\ref{mono2}, indicates that the relationship adapted unaltered from the polydisperse case is a good predictor of coexistence pressure for the monodisperse case as well. This is a clear indication that, as long as fractionation effects do not come into play, the master curve is overall independent of distribution width.
This behavior can be readily linked to the  plastic nature of the crystals formed by aspherical monodisperse systems~\cite{disorder1} for comparable values of $q$ and $A$, i.e.
each particle can freely rotate around their lattice sites, and can be understood qualitatively by realizing  
that any two identical neighboring particles will typically be misoriented and face each other with  a random region of their respective surfaces. 
As a consequence, their mutual interaction becomes  hardly distinguishable, on average, from that of two particles with different shape, thus explaining why monodisperse and polydisperse 
systems may indeed present analogous equilibrium properties. Note, however, the deviations from the predicted pattern 
 at  the extreme ends of $Aq$,  corresponding to particle shapes that have values of $q$ significantly 
different from the average $q$ of the distribution. The deviation at large $q$ is expected as the analogous behavior is observed for polydisperse systems;
as particles become more anisotropic their rotational degrees of freedom are reduced until they perfectly align in the $q\rightarrow 1$ (rod) limit, making 
the averaged random-shape argument described above inappropriate.
A  bit more surprising is the deviation for very small values of $q$, for which we find that the polydisperse system with $N_b=12$ nicely follows the master curve.
To rationalize this behavior one has to realize that for small values of $N_b$, when $q\rightarrow 0$, particles tend to acquire rather symmetric and specific geometries, such as platonic solids, which 
also require orientational ordering  to tile the space as soon as $A$ becomes sufficiently large. These specific particle shapes dominate the shape space for $q\sim 0$ and small values of $N_b$,
and, as discussed in~\cite{disorder1},  lead to deviations from the inverse power law behavior.  

Notice that no coexistence pressures above $p^*(A,q) \simeq 21$ are reported. This is interesting because this value is close to the fluid pressure of a system of hard spheres
at the glass transition density~\cite{speedy}, $p_G = 22.6$, and  it is reasonable to assume that $p_G$ sets an upper bound to the 
largest accessible fluid coexistence pressure for {hard} aspherical particles obtained with direct fluid/solid sampling. In fact, this method  
is clearly susceptible to anomalies in system kinetics, thus making coexistence measurements at pressures larger than the one here reported 
quite cumbersome and somewhat unreliable.
Nevertheless, this provides a way of rationalizing the empirical data, because if we assume that no aspherical hard particle will easily crystallize above $p_G$, we can readily link the pressure master curve $p^*(A,q)$
to the inverse power-law behavior obtained for monodisperse aspherical systems. By plugging $p_G$ into $p^* (A,q)$, we obtain the crystallizability boundary
$Aq\simeq 0.011$; this equation is plotted on Figure~\ref{mono1} along with the original line presented in~\cite{disorder1}.

Although it is not a perfect division, likely due to many factors (including finite-size effects), this simple prediction provides a good dividing line which would predict the {packing into fcc crystals} of about 80\% of all systems tested.
Given the uncertainties associated to the data points in~\cite{disorder1},  we believe this is a remarkably good rate of success, and provides an elegant physical reason for the inverse power law relationship found therein.

\section{CONCLUSIONS}

To conclude, our data provide a simple way of rationalizing the crystallizability limit observed for systems of monodisperse {hard} aspherical particles 
and generalize the results to include systems containing non-identical {hard} aspherical particles.
Our results show  that the coexistence pressure of systems of monodisperse and shape-polydisperse {hard} particles {with fcc crystals} is remarkably well-described, within the limits described in the text,   
 by a simple linear relation in $Aq$, $\frac{p^* - p^*_{HS}}{p^*_{HS}} = \alpha Aq$,  and indicate precise limits for the manufacture of 
nano-components expected to crystallize. Our study is by no means exhaustive of  all possible shapes, but it represents a first major step towards a statistical understanding of the role of 
shape disorder in particle organization. Our theoretical predictions should be readily testable experimentally.

\section*{Acknowledgments}
This work was supported by the American Chemical Society under  PRF grant No. 50221-DNI6.

\end{document}